\documentclass[journal]{IEEEtran}  

\usepackage[utf8]{inputenc}  %
\usepackage{tikz}
\usetikzlibrary{shapes.geometric, arrows.meta, positioning, fit}

\usepackage{xcolor}

\usepackage{tabularx}
\usepackage{booktabs}
\usepackage{ragged2e} 
\usepackage{longtable}

\usepackage{graphicx} 
\usepackage{subcaption}
\usepackage{amsmath}
\usepackage{amssymb}
\usepackage{cite}
\usepackage{hyperref}
\usepackage{color}
\usepackage{multirow}

\usepackage[ruled,vlined]{algorithm2e}
\usepackage{amssymb}

\begin{document}

\title{Behavior-Centric Malware Classification with Fine-Grained Malicious Logic Localization}
		\author{Prakriti Baral,
            Zhuoyun~Qian,
            Hailu~Xu,~\IEEEmembership{Member,~IEEE,}
           Fangtian Zhong,~\IEEEmembership{Member,~IEEE,}
            \thanks{Prakriti Baral is with the Gianforte School of Computing, Montana State University, Bozeman, MT 59717, USA, E-mail: prakritibaral22@gmail.com}
               
			\thanks{Fangtian~Zhong (Corresponding author) and Zhuoyun~Qian are with the College of Computer Science and Technology,
				 Shandong University, Qingdao,
				Shandong 266237, China. E-mail: fzhong@sdu.edu.cn and 202699900021@email.sdu.edu.cn}

            	\thanks{Hailu~Xu is with the Department of Computer Engineering and Computer Science ,
				 California State University, Long Beach, Long Beach, CA 90840, USA, E-mail: hailu.xu@csulb.edu}
        }
	\markboth{Latex}%
	{Shell \MakeLowercase{\textit{et al.}}: Bare Demo of IEEEtran.cls for IEEE Journals}
	%



	\maketitle
\begin{abstract}
Effective malware analysis requires understanding not only whether a program is malicious, but also which behaviors it exhibits and where those behaviors originate in the code. Existing machine-learning–based malware detectors largely operate as black boxes, providing limited insight into the malicious logic responsible for their decisions. This paper addresses malicious behavior localization and classification at the basic-block level. We propose a behavior-centric analysis framework that decomposes malware samples into behaviors and systematically links these behaviors to their originating code regions. Using context-sensitive backward slicing from security-relevant system API calls, we reconstruct control- and data-dependency chains and represent each behavior as a structured graph of related basic blocks. A Transformer-based model captures instruction-level semantics, while a Graph Neural Network models structural dependencies within behavior graphs. The resulting representations are fused to enable accurate and interpretable classification, with attention-based attribution identifying code regions responsible for malicious behaviors. We evaluate our approach using standard classification metrics and a behavior coverage metric that measures the detection of manually labeled malicious behaviors. Our results demonstrate that the proposed framework achieves accurate malware classification while providing fine-grained, behavior-aware localization of malicious logic.

\end{abstract}

\section{INTRODUCTION}

Malware detection has gradually shifted from signature-based techniques to learning-based approaches that model program structure and behavior. This transition is driven by the rapid growth in the volume, diversity, and sophistication of malware, which renders fixed signatures increasingly ineffective. Recent studies \cite{mohammadian2024explainable, bai2009detecting, fan2018android, mitra2023survey} leverage intermedaite representations such as control-flow graphs (CFGs) and function-call graphs (FCGs) to preserve execution semantics and information flow, thereby improving robustness to superficial code variations that do not alter program structure. To counter the increasing diversity of malware, researchers have developed a range of analysis-based detection techniques that differ mainly in feature extraction methods. Broadly, malware analysis approaches can be categorized as static, dynamic, or hybrid. Static methods \cite{baldangombo2013static, christodorescu2003static, alam2024revisiting, feriedooni2016anastasia} analyze binaries without execution, extracting features from byte sequences, opcodes, or API usage. Dynamic analysis techniques \cite{anderson2011graph, bhatia2017malware, jeon2020dynamic} observe runtime behaviors such as system calls, file operations, and network activity in sandboxed environments. Hybrid approaches \cite{shijo2015integrated, sugunan2018static, damodaran2017comparison, ijaz2019static} attempt to combine static and dynamic features to improve coverage.

Recent work \cite{lo2022graph, malhotra2023comparison} demonstrates that graph-based learning, particularly with Graph Neural Networks (GNNs), improves malware detection by explicitly modeling structural relationships such as control and data dependencies. In parallel, sequence models originally developed for natural language processing have been adapted to binary analysis. Transformer-based models \cite{almakayel2024deep, stein2024transformer, saracino2023graph} learn semantic representations of instructions and basic blocks without manual feature engineering and are effective at capturing long-range dependencies within code. These representations have been successfully applied to tasks such as function similarity detection, signature reconstruction, and binary classification.

Despite recent progress, most existing malware classification methods operate at the program or graph level and offer limited interpretability. They typically assign a label to an entire binary without identifying the specific code regions that implement malicious behavior, making it difficult for analysts to understand why a sample is classified as malicious and which malicious category it belongs to. More importantly, malware categories are fundamentally defined by their malicious behaviors. A meaningful classification approach should therefore focus on locating the code sequences that realize these behaviors and use them as the basis for category prediction. Existing methods generally do not provide this capability, as they rely primarily on global structural or statistical patterns rather than behavior-relevant code evidence. This limitation also reduces robustness against adversarial manipulation. Malware authors can modify instructions or alter control flow while preserving the underlying behavior, thereby changing the program’s overall appearance without affecting its malicious functionality. Such transformations can be generated automatically using techniques such as reinforcement learning \cite{zhong2022reinforcement} and GANs \cite{zhong2023malfox}. As a result, models that depend on whole-program patterns may be misled, whereas behavior-oriented analysis is more likely to remain stable because it focuses on the code directly responsible for malicious actions.

To address these limitations, we propose a behavior-centric malware classification framework that classifies malware according to the code sequences that implement malicious behavior and localizes the malicious logic responsible for each behavioral category. A central challenge in this setting is how to define behavior in a way that is both semantically meaningful and operationally tractable. In this work, we use security-relevant system API calls as analysis anchors rather than as a complete definition of malicious behavior. This choice is motivated by the observation that many high-impact malicious actions eventually materialize as security-sensitive interactions with the operating system, such as manipulating files, processes, memory, the registry, network channels, or privilege-related resources. These API invocations therefore provide concrete and observable entry points for isolating behavior-relevant code. Starting from these anchors, we construct control-flow graphs and perform context-sensitive backward slicing to trace the code regions that contribute to each behavior back to their entry points. The slicing process preserves interprocedural semantics through explicit call-return matching with a stack-based traversal. Each recovered behavior is then represented as a behavior graph, whose nodes correspond to basic blocks and whose edges encode control dependencies. Because the presence of a security-relevant API call alone does not imply malicious intent, we further manually annotate the basic blocks that genuinely contribute to harmful behavior, yielding fine-grained supervision for both behavior-level classification and localization. We do not claim that all malicious behavior must appear in this form; rather, we focus on the broad and practically important class of behaviors that manifest through security-relevant system interactions, for which our formulation offers a precise and reproducible analysis basis.

Each basic block’s instruction sequence is embedded using a Transformer model to capture semantic information, while a GNN models the structural dependencies within behavior graphs. This design is motivated by the nature of the malicious behaviors extracted in our framework. Specifically, each extracted behavior consists of a set of behavior-relevant basic blocks together with the control relationships that connect them. As a result, an effective model must capture both the semantics of individual code sequences and the structural organization of the overall behavior. The Transformer is used to encode the instruction sequence of each basic block because it can model contextual dependencies within code and produce semantically rich block-level representations. The GNN is then applied over the behavior graph to capture how these blocks interact through control dependencies, which is critical because malicious behavior is often determined not by a single block in isolation but by the coordinated execution of multiple related blocks. We train these two representations independently and later fuse them to integrate sequential and structural views of program behavior. Attention mechanisms highlight the behaviors that contribute most to classification decisions, and we introduce a behavior-coverage metric that quantifies the overlap between high-attention behaviors and manually labeled malicious regions, enabling measurable and interpretable localization. Our contributions can be summarized as follows:
\begin{itemize}
    \item We propose a unified framework that classifies malware samples while precisely localizing malicious logic at the basic-block level.

    \item We design a hybrid Transformer–GNN architecture that captures instruction-level semantics and structural dependencies in a complementary and interpretable manner.

    \item We introduce a behavior-coverage metric that connects attention-based model outputs with manually labeled malicious behaviors, enabling quantitative evaluation of localization quality.

    \item We evaluate our framework across multiple malware categories and demonstrate improved precision, recall, and interpretability compared to existing approaches

\end{itemize}

\section{Background and Related Work}
Malware detection and classification research has evolved 
along three primary dimensions defined by how features are 
extracted from malware programs: static analysis, dynamic 
analysis, and hybrid analysis. Each category has produced 
significant advancement, yet also carries inherent limitations 
that motivate our behavior-centric approach.

\subsection{Static Analysis-Based Approaches}
Static analysis examines malware without executing it, enabling efficient extraction of features from executable structure~\cite{christodorescu2003static}, control-flow graphs and API sequences~\cite{singh2017api}, and PE headers, imports, and API calls~\cite{baldangombo2013static}. Later work used graph representations for malware classification, including Android call graphs and function-call subgraphs~\cite{lo2022graph,malhotra2023comparison,fan2018android}. Transformer-based approaches have analyzed static feature vectors~\cite{almakayel2024deep}, manifest data~\cite{rahali2021malbert}, and API call sequences~\cite{saracino2023graph}. Related work has also applied Transformers to network-packet payloads~\cite{stein2024transformer}.Visual methods convert binary content into images or self-similarity descriptors for CNN-based classification~\cite{zhong2024enhancing,kalash2018cnn,naitabdesselam2020apps}. VisUnpac adds unpacking and contrast enhancement to improve the analysis of packed samples~\cite{zhong2025visualpatterns}.

A limitation of static classification is that features may reflect code that is present but never executed. A classifier could therefore rely on an unreachable routine without establishing that the routine contributes to malicious behavior. CFGExplainer~\cite{herath2022cfgexplainer} 
attempts to address this by targeting interpretability 
in GNN-based malware classifiers, using interconnected 
neural networks to score node importance within 
control-flow graphs. However, its explanation quality is evaluated by fidelity, measuring whether a pruned subgraph fed back into the same classifier reproduces its original prediction. It captures agreement with the classifier's own decision rather than alignment with independently verified malicious behavior.

\subsection{Dynamic Analysis-Based Approaches}
Dynamic analysis observes malware behavior at runtime, 
capturing system calls, API invocations, file 
operations, registry modifications, and network activity 
in sandboxed environments. This approach is inherently 
more robust to syntactic obfuscation because behavioral 
patterns tend to remain consistent even as code is 
transformed. Zolkipli and 
Jantan~\cite{zolkipli2011behavior} proposed a host-based dynamic framework that executes samples inside a Windows virtual machine, obtaining behavioral characteristics through runtime analysis and resource monitoring, and classifies malware sharing these characteristics using a knowledge-based technique. Hansen et~al.~\cite{hansen2016behavioral} introduced a scalable dynamic framework that executes samples inside a modified, distributed Cuckoo sandbox running on parallel virtual machines, recording the sequence and frequency of API calls and their input arguments during execution; these traces, combined with signature-based features drawn from AV-vendor malware encyclopedias, are fed into a Random Forests classifier for both binary detection and multi-class family classification, with Information Gain Ratio applied for feature selection.  
Du et~al.~\cite{du2019classified} proposed Magpie, 
which links API traces to kernel object parameters 
across seven functional categories,  represents these 
relationships as Classified Behavior Graphs, and uses a 
Support Vector Machine for classification. Anderson 
et~al.~\cite{anderson2011graph} demonstrated 
graph-based malware detection through dynamic analysis, 
showing that runtime behavioral traces captured in graph 
form improve detection of previously unseen malware 
variants. Dini et~al.~\cite{dini2012madam} developed 
MADAM, a multi-level host-based framework  
that simultaneously monitors security at the kernel, 
application, user, and package levels, correlating 
dynamic features including system calls and API call 
patterns with static package metadata using parallel 
KNN classifiers.

While dynamic approaches capture genuine execution 
behavior, they are constrained by typically short 
sandbox windows, which means important behaviors that 
trigger only under specific conditions may never be 
observed. Furthermore, these methods largely treat 
behavior as a global feature vector, making it 
difficult to connect predictions back to specific code 
regions that implement the malicious intent.

\subsection{Hybrid Analysis-Based Approaches}
Hybrid approaches attempt to combine static and dynamic 
features to gain broader coverage and overcome the 
individual limitations of each method. Shijo and 
Salim~\cite{shijo2015integrated} demonstrated that 
integrating static and dynamic features yields improved 
detection compared to either approach alone. Damodaran 
et~al.~\cite{damodaran2017comparison} provided a 
systematic comparison of static, dynamic, and hybrid 
strategies, finding that hybrid methods improve coverage 
but inherit the computational overhead and evasion 
challenges of both constituent approaches. Ijaz 
et~al.~\cite{ijaz2019static} further validated that 
machine learning applied to combined static and dynamic 
features outperforms single-modality approaches across 
diverse malware families. Xu and 
Chen~\cite{xu2023familial} introduced Behavior Trees 
mined from API call sequences to capture semantics 
within control structures such as loops and parallel 
executions, converting them into binary relation graphs 
processed by heterogeneous GNNs including simple-HGN 
and a modified RGCN for malware family classification.

\subsection{Summary}
Prior malware-classification research has made strong 
progress using static features, dynamic traces, and 
hybrid approaches, but most methods still fall short 
in two key ways: (1) they classify without reliably 
explaining where the malicious logic is, and (2) they 
provide limited, non-grounded interpretability. Static 
methods improve structural modeling through graph-based 
and visualization-based representations, yet they 
operate on the binary as a whole and do not connect 
classification decisions back to the specific code 
regions that implement malicious intent. 
Behavior-trace and API-feature systems such 
as~\cite{hansen2016behavioral} can scale, but they 
largely treat behavior as a global feature vector, and 
their conclusions depend on what is executed within 
typically short sandbox windows, so important behaviors may never 
appear, and ``why'' the model predicted a family 
remains unclear. Hybrid approaches improve coverage 
by combining static and dynamic signals, but they 
inherit the limitations of both: evasion-prone sandbox 
windows and no mechanism for localizing malicious logic 
within code. Sequence and Transformer-based 
approaches~\cite{rahali2021malbert} similarly focus on 
detection and categorization accuracy from API 
sequences, but they do not connect predictions back to 
specific basic blocks or code regions that implement 
the malicious intent. Finally, although 
CFGExplainer~\cite{herath2022cfgexplainer} explicitly targets explainability, its post-hoc, fidelity-based explanations lack ground truth. They assume that wherever the model focuses is automatically "important," so if a GNN misclassifies malware based on a benign compiler artifact, CFGExplainer will highlight that artifact as the explanation.

To overcome these gaps, our framework is 
behavior-centric and supervision-grounded: we decompose 
each sample into distinct behaviors via 
context-sensitive backward slicing from API sinks to 
entry points, represent each behavior as behavior 
graphs, and then manually assign maliciousness 
supervision (scores/labels) to the behaviors which 
contains a list of basic blocks that truly implement 
harm addressing the fact that not every API call 
indicates malicious intent. We learn dual 
representations, a Transformer for 
instruction$\rightarrow$BB semantics and a GNN for 
behavior-subgraph structure, and evaluate them 
separately and in fusion to test whether semantics, 
structure, or both best capture family-specific motifs. 
We then use Transformer attention, not just as a 
black-box signal, but as an explainability mechanism 
that can be quantitatively validated by measuring how 
well high-attention regions overlap with manually 
labeled malicious behaviors (behavior-coverage), 
directly answering both ``why this family?'' and 
``where is the malicious logic?'', a capability that 
many existing classifiers lack.

\section{Design and Framework}



\subsection{Overview of Behavior-driven Malware Classification Framework} 
\begin{figure}[htbp]
    \centering
    \includegraphics[width=\columnwidth]{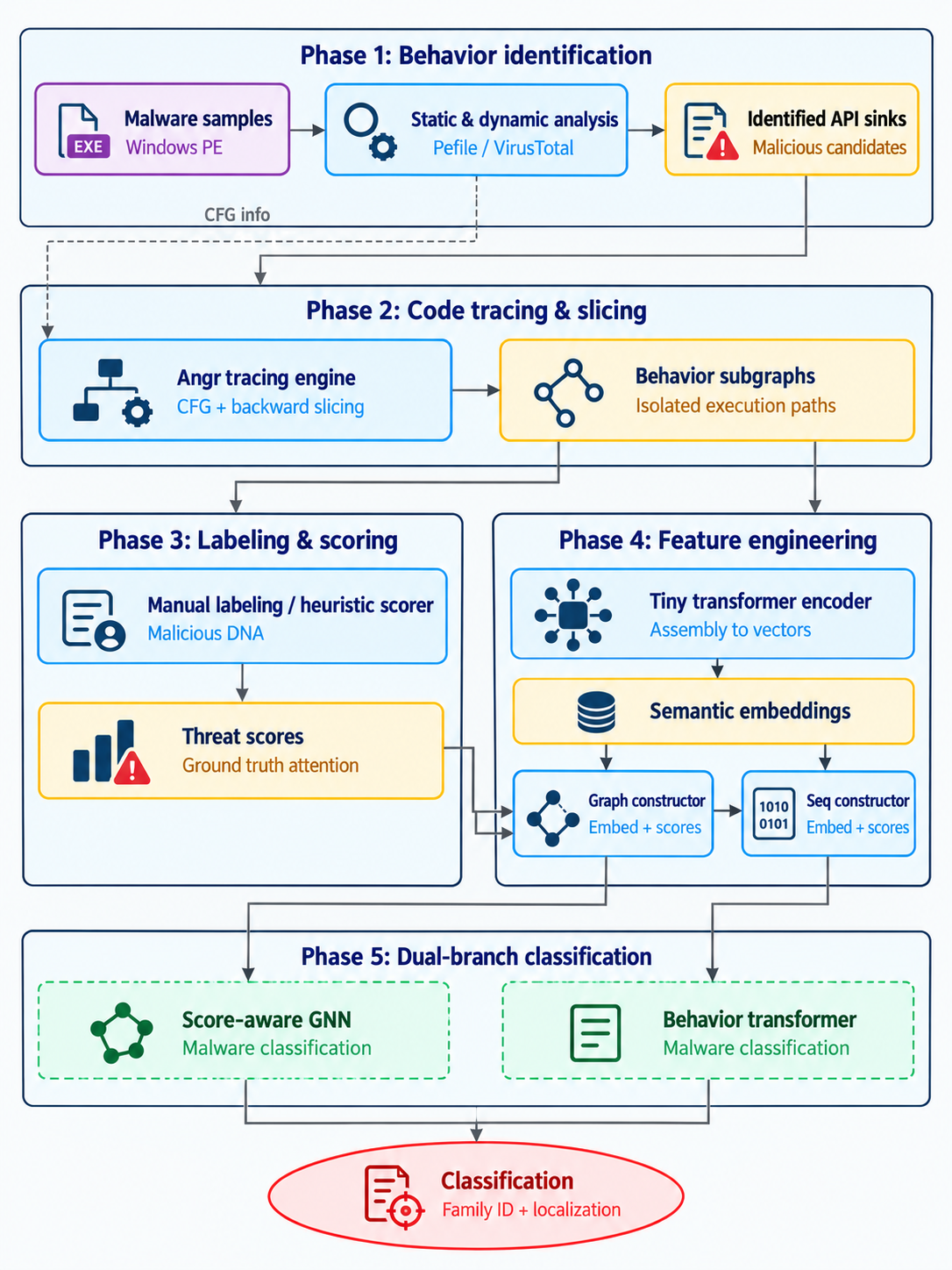}
    \caption{Overview of the behavior-driven malware classification framework.}
    \label{fig:framework_overview}
\end{figure}

The four primary objectives of our framework are: i) Accurate malware classification by modeling behavior traces ii) Robustness to structural and syntactic variation across variants: remain resilient to polymorphism/obfuscation and minor code changes by focusing on higher-level behavioral intent and execution context, iii) Discriminative, score-aware representations that reduce ambiguity between similar families and iv) Provide explainability and localization to avoid black-box decisions.
The framework, shown in Fig.~\ref{fig:framework_overview}, integrates several key stages to support behavior-centric malware classification and localization. The pipeline begins by ingesting raw Windows Portable Executable (PE) files. A static and dynamic analysis stage extracts function-level evidence and identifies API sinks calls to sensitive system functions that act as reliable indicators of suspicious intent. To avoid diluting the signal with benign boilerplate, we focus subsequent processing on code regions that contribute to these sink invocations.

Next, we construct a control-flow representation of the binary and perform backward tracing from each sink to recover the execution context that enables it. This yields isolated execution paths, represented as ordered traces of basic blocks (behavior subgraphs). Each subgraph forms a self-contained unit of execution logic, allowing the malware’s capabilities to be separated into independent behavioral components. We manually examine the extracted behaviors and assign maliciousness scores based on recurring family-specific patterns. A heuristic scorer then applies a set of Malicious DNA rules to these annotations, producing two related but distinct measures. The first is a maliciousness score: a raw, unbounded integer computed at the basic-block level from weighted API roles plus bonuses for suspicious patterns. Because this score is sample-local, it is not comparable across samples — a score of 50 in one sample carries no fixed meaning relative to a score of 50 in another.
The second is a normalized threat score: a float in [0,1] computed at the behavior-path level. This is obtained by aggregating the raw basic-block scores along each path, min-max normalizing across all paths within a sample, and then thresholding to produce a binary label. Unlike the raw score, this normalized value is comparable within a sample and serves as an explicit supervision signal — one that distinguishes auxiliary code from behavior-critical regions and provides a reference point for interpretability. 

After scoring, the framework converts behavior subgraphs into two complementary learning views. First, basic-block assembly is encoded into dense semantic embeddings using a compact Transformer-based encoder. Second, we construct both a structural view (capturing relationships among basic blocks) and a sequence view (capturing the ordered narrative of execution). Both views start from the same per-basic-block embedding, which already has the block's threat score appended to it. For the structural view, each basic block keeps its own embedding as a separate node, and consecutive blocks along a behavior's execution path are connected by directed edges, forming a small chain-like graph per behavior. For the sequence view, the per-block embeddings belonging to one behavior are instead mean-pooled into a single vector representing that whole behavior, and all such behavior vectors from a sample are stacked in their extracted order into one ordered sequence.

Before either view is passed into its respective model, both branches apply a shared score-aware rescaling step, each governed by its own learnable scalar weight. In the structural view, every node's feature vector is rescaled in proportion to its threat score, so that blocks carrying higher threat scores are amplified relative to low-score ones; individual behavior graphs are then combined into a single batched graph object, with feature dimensions padded or aligned as needed, before being processed by the graph convolution layers. In the sequence view, each pooled behavior vector is first linearly projected into the model's hidden dimension and then rescaled by the same threat-score-driven mechanism before entering the Transformer encoder; since samples contain varying numbers of behaviors, sequences are padded to a common length within each batch, with a mask ensuring padded positions are excluded from self-attention. In both branches, then, the threat score is not a passive input feature but an active signal that reshapes the representation's magnitude ahead of the core model layers, biasing both the GNN and the Behavior Transformer toward behavior-critical evidence.

We evaluate the predictions from each branch independently and also fuse their outputs at the decision level to obtain the final malware family classification. This fusion happens per sample rather than per individual behavior graph. The Behavior Transformer naturally produces a single prediction vector per sample, since its input already spans the sample's full behavior sequence. The GNN, by contrast, produces one prediction per behavior graph, and since a sample may contain several such graphs, these are collapsed into a single sample-level vector by taking the elementwise maximum across all of the sample's graph-level predictions — allowing the strongest piece of structural evidence to drive the GNN's contribution. The two resulting sample-level vectors are then combined through weighted linear interpolation, governed by a single scalar weight selected by grid search over the validation split; the final family prediction is the argmax of this fused vector. This is a straightforward late, decision-level fusion rather than a jointly learned or attention-based one: the two branches are trained independently, and only their output logits are combined afterward using one tuned interpolation weight.

In addition to classification, the framework supports explainability by comparing expert-assigned threat scores with the model's learned importance signals, highlighting the behaviors and basic blocks most responsible for each decision.

\subsection{The Components of the Behavior-driven Malware Classification Framework}

\subsubsection{Behavior Identification and Subgraph Extraction}

The first stage of our framework converts a raw Windows PE executable into a set of isolated behavior subgraphs. The motivation is practical: real binaries contain large volumes of compiler/runtime scaffolding and benign library glue that are expensive to process and often weakly related to the malicious objective. 

The pipeline begins with static parsing of the PE structure to obtain an initial capability view of the sample. Imported functions are extracted from the Import Address Table, which provides a stable list of OS-level dependencies. To reduce over-reliance on explicit imports, particularly in samples where the IAT is incomplete, this capability view is augmented with additional static signals. String evidence extracted directly from the binary is used to capture API and DLL references. These references may be resolved dynamically at runtime, while a control-flow graph (CFG) is generated using angr to provide a disassembled, interprocedural view of basic blocks and call relationships. Through CFG construction, call instructions can be resolved to import stubs, external functions, or inferred call targets , enabling API usage to be localized beyond a naive import listing.

Because different malware families emphasize different tactics and packing strategies, this stage is further informed by triage evidence from sandbox reports of VirusTotal \cite{sood2025virustotal}, which is used to prioritize security-sensitive APIs as high-confidence sinks. To move from raw import strings to behavioral meaning, each API is mapped into a coarse taxonomy of intent categories (process/thread control, memory management, file activity, registry modification, UI/clipboard/GDI activity, synchronization/TLS, and networking). This mapping does not assume that every imported function is executed; rather, it establishes a principled candidate set of security-relevant actions that can be localized and validated in later stages. Not every call target recovered at this stage resolves to a named import — some correspond to internal calls that cannot be tied to a known API, and these are tracked separately rather than folded into the taxonomy, with their proportion varying by family and packing strategy. The output of this step is a curated sink vocabulary: APIs that are both security-sensitive and diagnostically useful for subsequent behavior reconstruction.

\subsubsection{Behavior Trace Reconstruction via CFG-Guided Backtracing} 
After defining the sink set, the framework localizes where these sinks occur in the code. We achieve this through static control-flow recovery using the angr\cite{shoshitaishvili2016sok}
 binary analysis framework. Specifically, we construct a CFGFast representation of the program. CFGFast yields a graph over basic blocks and functions, enabling us to reason about code structure. Using this CFG, each candidate sink is resolved to one or more concrete invocation locations by searching for sink-target nodes and then recovering their callsite basic blocks. Rather than treating the sink API node itself as the behavioral anchor, we associate the sink with the basic block that contains the call instruction, since this block is the most precise static location that can later be traced, embedded, and interpreted. Each localized callsite is recorded with its basic-block address, enclosing function context, and a semantic tag derived from the sink taxonomy, which pairs the coarse intent category established above (e.g., \emph{registry}) with a more specific behavior label describing the concrete action performed (e.g., writing a value). For example, a call to the registry API \texttt{RegSetValueExW} is localized to the basic block containing the call instruction, within its enclosing function, and tagged with intent category \emph{registry} and behavior label \emph{Reg Set}. In practice, multiple sink calls may appear within a single basic block; therefore, callsite evidence is consolidated at the basic-block level to avoid duplication and to produce a stable set of behavior anchors. The output of this step is a structured mapping of sink to callsite basic block along with per-callsite metadata. For the example above, this takes the form {basic-block address, enclosing function, category: \emph{registry}, behavior: \emph{Reg Set}}, which serves as the starting point for targeted slicing.

Once sink callsites are localized, the framework reconstructs the execution context that enables each sink. This is implemented as targeted backward slicing over the CFG. Using the CFGFast output, we build lightweight artifacts once per binary, most importantly a predecessor map that associates every basic block in the program's CFG — not only sink-adjacent blocks — with the set of blocks that can transfer control into it; this single whole-program map is then reused across all sink callsites. This reverse adjacency structure makes backward traversal efficient and avoids repeatedly querying the analysis engine during slicing.

For each sink callsite, slicing begins at the callsite basic block and proceeds backward through predecessor blocks, exploring multiple predecessor routes in parallel to recover plausible entry-to-sink chains; when a basic block is reachable via more than one predecessor route, each route is preserved as its own distinct chain rather than merged, so the same basic block may appear in several different chains without them being conflated. Because malware binaries may contain large control-flow regions and indirect branching, naive backward exploration can become intractable. To keep the process scalable, the slicing procedure bounds each callsite's search by a maximum traversal depth, a cap on predecessor expansions, and a per-callsite timeout, and retains a fixed number of raw candidate chains; after removing exact duplicates, the remaining chains are ranked by length, and only the shortest — the most direct entry-to-sink chains — are retained as the representative paths for that sink. 

Entry points are treated broadly, not only the program entry but also recovered function-entry blocks and CFG source nodes — basic blocks with no incoming edges in the recovered control-flow graph, typically reached only through indirect or dynamically-resolved control flow that static recovery could not connect — since malware often dispatches behavior from multiple routines.

The result of this process is an isolated behavior subgraph, represented as an ordered trace of basic blocks from an entry context to the sink callsite. Each subgraph is stored with its sink metadata (API identity and category), the callsite identifier, and the basic-block sequence that forms the extracted execution path. Conceptually, these subgraphs serve as the framework’s behavior units: they separate a monolithic program into multiple independent components, each corresponding to a localized capability (e.g., a network connection routine, a persistence-related registry update, or a process-manipulation sequence). These behavior subgraphs become the canonical inputs for the subsequent scoring and feature engineering stages.

\subsubsection{Labeling and Scoring}
A key design choice in our framework is the incorporation of manual behavior inspection. After extracting behavior subgraphs, we perform manual inspection to identify which subgraphs correspond to genuinely malicious objectives for a given family and which reflect benign scaffolding. This manual step produces two artifacts for the remainder of the pipeline: (i) a family-specific understanding of recurring malicious patterns (the “Malicious DNA” reference) and (ii) a curated set of behaviors that should receive higher emphasis during scoring and interpretation. These manual identifications do not replace learning; instead, they provide a grounded behavioral prior that is used later to assign threat scores and to evaluate explainability by comparing manual judgments to model-derived importance.

After behavior subgraphs are extracted, we manually inspect representative subgraphs across families to determine which subgraphs capture malicious core logic and which represent benign logic,compiler/runtime initialization, or generic helper code. This step is crucial because static slicing can legitimately extract both harmful and non-harmful contexts around sinks, for example a API usage that appears sensitive in isolation but is part of benign configuration or error handling. During manual review, subgraphs are interpreted at the level of behavior intent rather than raw instruction sequences. Different versions of malware belonging to the same family often share reused code \cite{calleja2019malsource}. We conducted review of representative subgraphs across the dataset to identify recurring patterns or "Malicious DNA" that characterize specific families. This involves examining combinations of sink categories, call patterns, and contextual setup blocks. Based on these observations from around 50 samples of each family, we defined a set of heuristic rules to assign threat scores to subgraphs. Each behavior receives a real-valued threat score in the range [0, 1], where 0 indicates negligible malicious relevance and 1 indicates high-confidence malicious relevance. A value such as 0.5 therefore indicates that a behavior's evidence sits midway between the weakest and strongest behaviors identified within that particular sample, rather than a fixed universal severity level. This manual identification stage directly supports two downstream goals.  First, it enables principled threat scoring, where subgraphs and/or their constituent basic blocks receive severity weights aligned with observed family patterns. Second, it anchors explainability: the model explanations can be validated by comparing what the model highlights against what are manually identified as behavior-critical. 

\subsubsection{Feature Engineering}
Following behavior extraction and threat scoring, the framework transforms raw basic-block code and heuristic scores into mathematical representations suitable for deep learning. Raw assembly instructions are highly variable across compiler versions, optimization levels, and packing strategies. Superficial differences such as register allocation, memory offsets, or immediate constants can change significantly without altering program logic. To mitigate this variability, we apply a normalization and tokenization process to the basic blocks extracted during behavior tracing.

For each behavior-relevant basic block, disassembled instruction streams are converted into a normalized token sequence. Operand-specific information is abstracted to reduce noise: memory addresses are replaced with generic memory tokens, registers are mapped to abstract register symbols, and large immediate values or string references are masked with constant tokens. For example, the instruction \texttt{mov word ptr [ebx + 4], 0xd7b2} is normalized to the token sequence \texttt{mov, MEM, BASE, DISP, IMM} — the mnemonic is preserved, the memory operand is abstracted to its structural role (a base register plus a displacement) rather than its concrete address, and the immediate value is masked entirely. Instruction mnemonics are preserved, as they encode the core computational intent of the block. This normalization strategy encourages the model to learn logical patterns instead of overfitting to sample-specific artifacts. The resulting instruction tokens are mapped into a domain-specific vocabulary constructed from the training corpus. Each basic block is represented as a fixed-length sequence of 64 integer token IDs, where shorter blocks are padded and longer blocks are truncated.

Traditional malware analysis often relies on one-hot encoding or simple word2vec embeddings, which fail to capture the complex context of assembly instructions. To overcome this, we employ a custom Tiny Transformer Encoder to generate semantic embeddings for each basic block. The encoder consists of an embedding layer, positional encodings, and multiple self-attention layers. It processes the sequence of token IDs from the normalization step, allowing the model to attend to the relationships between instructions. The Transformer outputs a dense, high-dimensional vector $E_{bb} \in \mathbb{R}^d$ for each basic block. 

Each basic block is associated with a normalized threat score $S_{bb}$~$\in$~[0,1], derived from the manual inspection. These scores reflect the degree to which a block aligns with known malicious behavior patterns. For each basic block, the semantic embedding is augmented with its corresponding threat score, producing a score-aware embedding ($E_{bb} \oplus S_{bb}$). This design ensures that downstream models are explicitly guided toward behavior-critical regions, while still retaining the flexibility to learn additional patterns from data.

Using score-aware embeddings, the framework constructs two complementary representations tailored to the strengths of the dual-branch architecture. For the GNN branch, each extracted behavior trace is converted into a graph  $G = (V, E)$. Nodes correspond to basic blocks, and edges encode the execution order along the recovered entry-to-sink trace. Each node feature vector includes the score-aware embedding, along with three auxiliary structural fields: a binary entry-point indicator (1 for the block that starts the chain, 0 otherwise), a binary sink-call flag (1 for the block containing the sink call that ends the chain, 0 otherwise), and a one-hot behavioral category encoding assigned to the sink block (e.g., \emph{file}, \emph{registry}), with all non-sink blocks defaulting to a \emph{misc} encoding. This representation allows the GNN to learn structural signatures of malicious behaviors while remaining explicitly informed by threat severity.

For the Behavior Transformer branch, each behavior trace is represented as an ordered sequence of basic blocks, arranged according to execution flow from entry context to sink. Each sequence element consists of the score-aware embedding together with two auxiliary labels: a binary malicious/benign label indicating whether that step was identified as malicious, and a categorical behavior-type label (e.g., \emph{file}, \emph{registry}, \emph{network}) indicating the kind of action it performs. This representation preserves the temporal narrative of the behavior, enabling the Transformer to model long-range dependencies and execution patterns.

We utilize a sharded merging strategy to consolidate the thousands of individual per-behavior graph objects generated in the previous step, which are produced one sample at a time and initially stored as a list of graphs per sample. This process aggregates the isolated graphs into a single PyTorch Geometric dataset. During this merger, a unique identifier (UID) of the form family/sample is attached to every graph object. This UID preserves the mapping between the graph structure and its source binary. Following the structural merger, we build an analogous corpus-wide dataset for the Transformer branch: for every sample, its scored behaviors and basic-block embeddings are collected and merged across the entire dataset into a single global behavior map and a corresponding merged embedding set, with each entry tagged by the same UID scheme used for the graphs. This module ingests that global behavior map and the merged embeddings, aligning them to produce a 3D tensor of shape $(N_{samples} \times L_{sequence} \times D_{embedding})$. This dataset explicitly encodes the time-series narrative of behaviors, ensuring that the Transformer model receives a consistent, ordered view of the malware's execution history corresponding exactly via the shared UID to the merged graphs.

\subsubsection{Malware Classification Using Transformer}

\begin{figure*}[t]
    \centering
    \includegraphics[width=\textwidth]{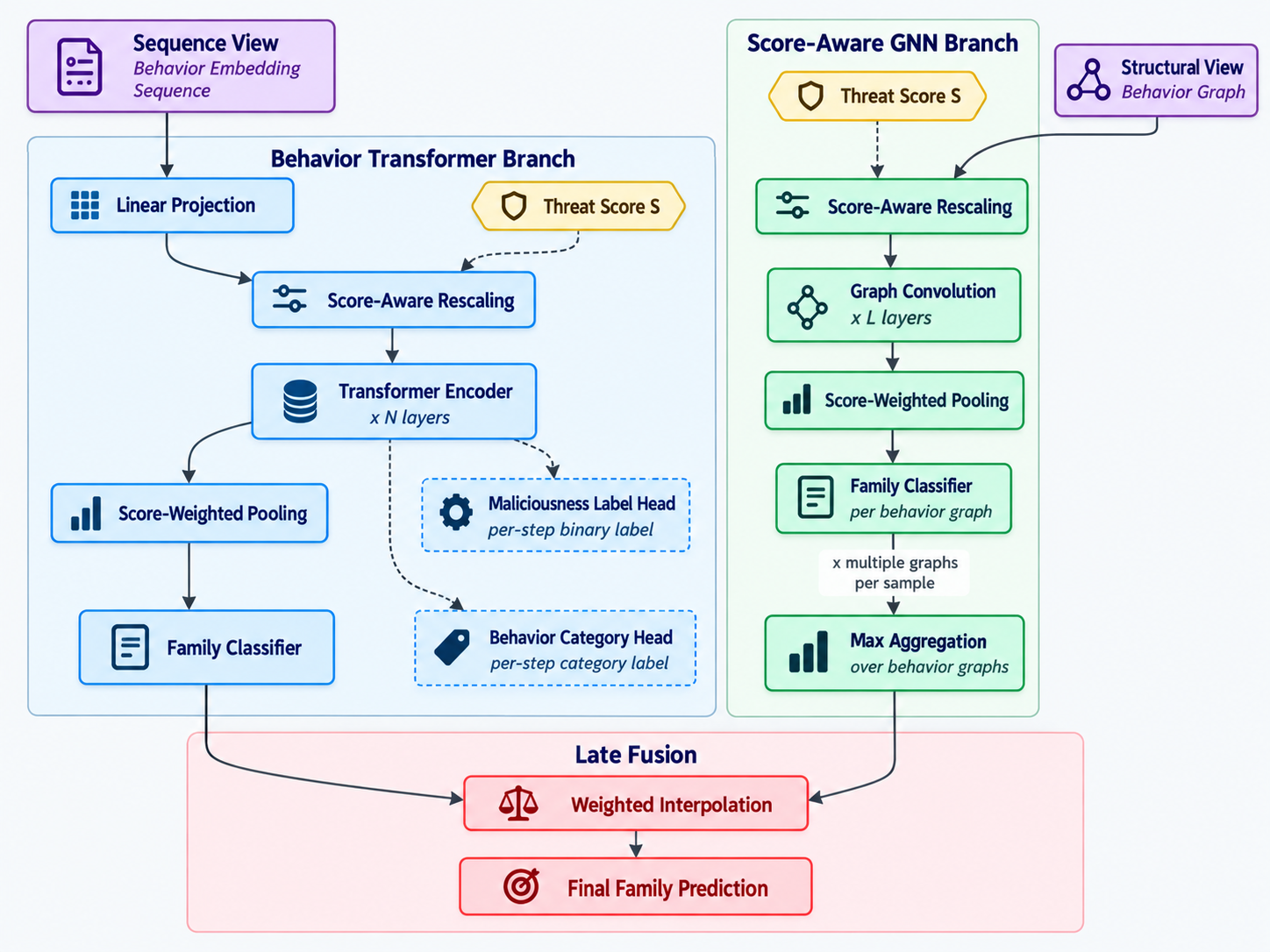}
    \caption{Detailed architecture of the Behavior Transformer branch,
    the score-aware GNN branch, and late fusion.}
    \label{fig:model_architecture}
\end{figure*}

Fig.~\ref{fig:model_architecture} shows the full architecture of the model. This subsection covers the Behavior Transformer branch on the left, and the GNN branch and fusion step are covered in the next two subsections. For input representation
in a given sample, we form a variable-length sequence of behavior
vectors
\[
\mathbf{X} = [x_1; x_2; \dots; x_M] \in \mathbb{R}^{M \times d_{in}},
\]
where $M$ is the number of extracted behaviors for that sample (the
sequence is padded per mini-batch). Each element $x_t \in \mathbb{R}^{d_{in}}$
is a behavior embedding produced during feature engineering — $d_{in}$
is the same behavior-embedding dimensionality denoted $D_{embedding}$
established during feature engineering — and each time step is associated
with a normalized threat score $S_t \in [0,1]$. In each mini-batch,
sequences are padded to the maximum length $M_{\max}$ and a binary mask
is maintained to distinguish valid steps from padding.

Inputs are first projected into a latent space of dimension $d_{model}$
using a linear projection:
\[
\mathbf{H}_0 = \mathbf{X} W_p \in \mathbb{R}^{M \times d_{model}},
\]
where $W_p \in \mathbb{R}^{d_{in} \times d_{model}}$; this projection
changes each step's dimensionality from $d_{in}$ to $d_{model}$. To
explicitly bias the representation toward high-threat behaviors, the
projected vectors are rescaled using the Malicious DNA scores and a
learnable scalar $\alpha$:
\[
\mathbf{H}_0 \leftarrow \mathbf{H}_0 \odot \left(1 + \alpha \, \mathbf{S}\right),
\quad \mathbf{S} \in \mathbb{R}^{M \times 1},
\]
where $\mathbf{S}$ contains the per-step threat scores, broadcast
across the $d_{model}$ feature dimension, and $\odot$ denotes
elementwise multiplication. The resulting sequence is then processed
by a stack of $N$ Transformer encoder layers using multi-head
self-attention and position-wise feed-forward networks, with
pre-normalization and residual connections. Padding positions are
excluded from attention via the key-padding mask.

Let $\mathbf{H} = [h_1; \dots; h_M] \in \mathbb{R}^{M \times d_{model}}$ denote
the final encoder outputs for the valid steps. We compute a program-level
representation using Malicious DNA score-weighted pooling:
\[
w_t = \frac{S_t}{\sum_{j=1}^{M} S_j},
\qquad
v_{seq} = \sum_{t=1}^{M} w_t h_t \in \mathbb{R}^{d_{model}}.
\]
If $\sum_{j=1}^{M} S_j = 0$ for a sample,
the pooling falls back to uniform averaging over valid (non-padding) steps.
The pooled vector $v_{seq}$ is passed to a linear classifier to produce
family logits and the predicted family label.

To encourage semantic robustness at the behavior level, the model is
trained with two step-level auxiliary objectives in addition to family
classification. For each behavior step $t$, the model predicts:
(i) a binary maliciousness label, derived from the same threat-score
threshold used to assign the Malicious DNA scores, and
(ii) a categorical behavior-type label, drawn from the VirusTotal-informed
category taxonomy used to build the sink vocabulary. Both auxiliary heads
operate on the per-step encoder outputs $h_t$. During training, auxiliary
losses are computed only on valid positions.

Importantly, the auxiliary losses are \emph{score-weighted}, using the same
weighting scheme as the pooling above. For each auxiliary head, per-step
cross-entropy terms $\mathrm{ce}_t$ are computed without reduction and then
combined as
\[
\mathcal{L}_{\text{aux}} = \sum_{t=1}^{M} \tilde{w}_t \, \mathrm{ce}_t,
\qquad
\tilde{w}_t = \frac{S_t}{\sum_{j=1}^{M} S_j},
\]
where the sum runs over valid (non-padding) steps only. When all valid
scores are zero, $\tilde{w}_t$ falls back to a uniform weight
$1/M_{\text{valid}}$ over those steps. This design ensures that
high-threat behaviors contribute more strongly to auxiliary supervision
without discarding low-score context.

The overall training loss is the sum of the primary family classification
loss and the score-weighted auxiliary losses:
\[
\mathcal{L}_{total}
=
\mathcal{L}_{family}
+
\lambda_{1}\mathcal{L}_{beh\_label}
+
\lambda_{2}\mathcal{L}_{beh\_cat}.
\]
The family loss $\mathcal{L}_{family}$ is optimized using cross-entropy
with label smoothing (smoothing factor $0.1$ during training):
\[
\mathcal{L}_{family} = -\sum_{c=1}^{K} y'_c \log \hat{p}_c,
\qquad
y'_c =
\begin{cases}
1-\epsilon, & c = y \\
\epsilon/(K-1), & c \neq y
\end{cases}
\]
where $K$ is the number of malware families, $y$ is the true family label,
$\hat{p} = \mathrm{softmax}(\text{family\_logits})$, and $\epsilon = 0.1$
is the smoothing factor. The default auxiliary weights are $\lambda_{1}=0.5$
and $\lambda_{2}=0.3$. Note that the per-step score-weighting described
above operates within each auxiliary loss term; $\lambda_1$ and $\lambda_2$
are separate, fixed scalar hyperparameters — not derived from threat
scores — that control the overall contribution of each auxiliary
objective to the total loss.

The model is implemented in PyTorch with $N=4$ encoder layers, hidden
dimension $d_{model}=256$, and $8$ attention heads. Dropout with rate
$0.4$ is applied within the Transformer layers and on the encoder
outputs. Optimization uses AdamW with learning rate $3\times10^{-4}$ and
weight decay $10^{-3}$. Training is performed for up to 30 epochs with
batch size 32. To mitigate class imbalance across malware families, we
use a family-balanced sampling strategy that approximately equalizes
family participation during training. Gradients are clipped with a
maximum norm of 1.0 for stability.

\subsubsection{Malware Classification Using GNN}

This branch performs malware family classification using a structural view of each extracted behavior subgraph. This branch is shown on the right in Fig. 2. Each behavior instance is
represented as a graph $G=(V,E)$, where each node $v\in V$ corresponds
to exactly one basic block, and edges $(u,v)\in E$ encode the recovered
execution order along the entry-to-sink trace. Each node carries a
feature vector $x_v\in\mathbb{R}^{d_{in}}$ produced during feature
engineering (semantic embedding augmented with structural metadata),
and includes an aligned Malicious DNA threat score $s_v\in[0,1]$ as one
of the feature dimensions, the same threat score assigned to that
specific basic block during Labeling and Scoring, looked up and attached
to its corresponding node. Before message passing, node features are
rescaled using the threat score and a learnable scalar $\alpha$ to bias
the model toward behavior-critical regions:
\[
x'_v = x_v \cdot (1 + \alpha s_v),
\]
where $\cdot$ denotes scalar multiplication of the feature vector $x_v$
by the per-node scalar $(1+\alpha s_v)$. This operation preserves the
full feature vector while amplifying nodes associated with higher
threat scores.

We apply a stack of $L$ graph convolution layers (GCN) to propagate
contextual information across the behavior subgraph. Let
$h_v^{(0)}=x'_v \in \mathbb{R}^{d_{in}}$; each layer performs
neighborhood aggregation followed by a nonlinearity:
\[
h_v^{(\ell+1)} = \sigma\!\left(\sum_{u\in\mathcal{N}(v)} \tilde{w}_{uv}\, h_u^{(\ell)} W_\ell\right),
\qquad h_v^{(\ell)} \in \mathbb{R}^{d_\ell},
\]
where $W_\ell \in \mathbb{R}^{d_\ell \times d_{model}}$ are learnable
weights (with $d_0 = d_{in}$ and $d_\ell = d_{model}$ for $\ell \geq 1$),
$\sigma(\cdot)$ is ReLU, and $\tilde{w}_{uv}$ are the GCN normalization
coefficients, computed as the fixed, precomputed structural weight
\[
\tilde{w}_{uv} = \frac{1}{\sqrt{\hat{d}_u \hat{d}_v}}, \qquad \hat{d}_i = 1 + \deg(i),
\]
with $\deg(i)$ the degree of node $i$ in the behavior subgraph and the
$+1$ accounting for an added self-loop. Note that edges themselves carry
no learnable parameters in this formulation — only the node feature
transforms $W_\ell$ are learned; edges determine which neighbors are
aggregated and are weighted only by the fixed coefficient $\tilde{w}_{uv}$
above. Dropout is applied after each layer to reduce overfitting. After
the final convolution layer, node embeddings are aggregated into a
single graph-level representation via Malicious DNA score-weighted
pooling:
\[
h_G=\frac{\sum_{v\in V} s_v\, h_v^{(L)}}{\sum_{v\in V} s_v + \varepsilon} \in \mathbb{R}^{d_{model}},
\]
where $\varepsilon$ is a small constant to avoid division by zero. The
pooled vector $h_G$ is passed through a linear classifier to produce
family logits, and the model is optimized using cross-entropy loss.

The model is implemented in PyTorch Geometric using a GCN backbone with
hidden dimension 256, $L=3$ convolution layers, and dropout 0.4.
Optimization uses Adam with learning rate $10^{-3}$ and weight decay
$10^{-4}$. Training is performed for up to 30 epochs with batch size 32.
Gradients are clipped with a maximum norm of 5.0 for stability.

\subsubsection{Malware Classification Using Late Fusion}

Late fusion combines the two branches at the decision level, after the Transformer and the GNN have each produced their own family predictions, a K-dimensional logit vector, where K is the number of malware families and each entry corresponds to one candidate family. This step is shown at the bottom of Fig. 2. Fusion is performed at the program (UID) level so that each malware binary receives exactly one final label. The Transformer already outputs a single logit vector $z_{tr} \in \mathbb{R}^K$ per UID. The GNN, however,
produces one logit vector $z_g \in \mathbb{R}^K$ per behavior graph,
since one binary yields multiple extracted behavior subgraphs. To
obtain a UID-level GNN prediction, we aggregate the logit vectors of
all graphs belonging to the same UID using an elementwise maximum,
\[
z_{gnn} = \max_{g \,\in\, \text{UID}} z_g \in \mathbb{R}^K,
\]
so the strongest structural evidence from any behavior can influence
the final decision, the dimensionality remains $K$ throughout this
step, only the values change. The Transformer logits $z_{tr}$ and
aggregated GNN logits $z_{gnn}$ are then combined through a weighted
interpolation controlled by a scalar $\beta \in [0,1]$, chosen on the
validation split:
\[
z_{fused} = \beta\, z_{gnn} + (1-\beta)\, z_{tr} \in \mathbb{R}^K.
\]
The final malware family prediction is $\arg\max_k (z_{fused})_k$,
while the UID alignment ensures that the sequential and structural
evidence being combined always refers to the same source binary.

\section{Evaluation Metrics}

To fully evaluate our models' performance, we use a dual-stage evaluation approach that measures the model's ability to locate malicious behaviors within code as well as the accuracy of malware family classification.

\subsection{Classification Performance Metrics}
To measure model performance across 43 malware families, we adopt four commonly used performance metrics, including Precision, Recall, F1, and Accuracy.



\subsection{Behavior Identification Metric}

Beyond family classification, we introduce a behavior-level metric to evaluate the model’s explainability and localization ability.
For a given sample, let $B_{GT}$ denote the set of behaviors labeled as malicious, and let $B_{TopK}$ denote the top-$K$ behaviors ranked by model contribution. Coverage is computed as:
\[
\text{Coverage} = \frac{|B_{GT} \cap B_{TopK}|}{|B_{GT}|}.
\]
This metric captures how effectively the model’s explanation
recovers manually identified malicious behaviors.

\subsection{Experiment Setup}
PyTorch v2.3.0 and angr v9.2.63 were used as the primary deep learning and static program analysis backends, respectively. The complete framework was implemented in Python 3.11. To support reproducibility, we fixed random seeds across all experiments and used consistent data partitions for training, validation, and evaluation. Model checkpoints, configuration parameters, and evaluation logs were saved for every run, enabling deterministic reruns under identical settings.

Computational efforts were performed on the Tempest High Performance Computing System, operated and supported by University Information Technology Research Cyberinfrastructure \cite{msu_tempest_rrid} (RRID:SCR\_026229) at Montana State University. Jobs were submitted through the Slurm workload manager. CPU-based runs used a typical allocation of 60 CPU cores and 200~GB RAM per job, which was sufficient for large-scale graph and tensor streaming workloads and ensured stable throughput during evaluation. 

\subsection{Dataset}


In this work, we rely on the malware dataset compiled by Zhong et al. \cite{zhong2025visualpatterns} and \cite{zhong2024enhancing}. The dataset used in this study was recollected and refined using aggregated reports from over eighty antivirus engines on VirusTotal \cite{zhong2024enhancing}, followed by label cleansing and verification using Cuckoo sandbox dynamic analysis \cite{cuckoo_docs_what} and manual inspection to ensure consistency.

\section{EVALUATION RESULTS}

We begin by evaluating the accuracy of our Transformer, GNN, and
late-fusion models on the dataset, followed by a comparison with three
additional methods that use different techniques. As shown in Table~I,
the Behavior Transformer achieves consistently strong performance, with
an average precision of 99.62\%, recall of 99.61\%, F1 score of 99.61\%,
and accuracy of 99.61\%. The score-aware GNN also performs competitively,
obtaining an accuracy of 99.00\%, precision of 98.75\%, recall of
99.04\%, and F1 score of 98.88\%. Finally, combining both branches via
late fusion yields the best overall results, achieving 99.87\% across
accuracy, precision, recall, and F1. These results indicate that the
sequential and structural views capture complementary signals: the
Transformer provides strong sequence-level discrimination, while the
GNN contributes additional structural evidence that improves the final
fused decision. Tables~II and III summarize the per-family
classification reports for the Transformer and GNN models, respectively.
Overall, both models achieve high precision and recall for the majority
of malware families, indicating stable discrimination across classes
rather than performance being driven by only a small subset. The
remaining errors are concentrated in families with similar behavior
patterns. For example, gandcrab\_trj shows the lowest precision among
the reported families (0.8571 for the Transformer, Table~II), well
below the 0.98--1.00 range achieved by the other families. This is
likely because both models rely on broad behavior-category signals
rather than fine-grained instruction-level differences, making families
with overlapping high-level tactics, such as similar registry, process,
or file-staging behaviors, harder to distinguish.

\begin{table}[htbp]
\centering
\caption{Accuracy of Different Models}
\label{tab:sample_table}
\begin{tabular}{|c|c|c|c|c|c|}
\hline
Models & Precision & Recall & F1 & Accuracy  \\
\hline
Transformers & 99.62 & 99.61  & 99.61 & 99.61   \\
\hline
GNN & 98.75  & 99.04 & 98.88 & 99.0   \\
\hline
Late Fusion & 99.87 & 99.87 & 99.87  & 99.87  \\
\hline
CFGExplainer \cite{herath2022cfgexplainer} & 72.2  & 71.8 & 70.24  &   83.7  \\
\hline
MalBERT \cite{rahali2021malbert} & 95.86  & 92.57  & 91.68 & 90.91 \\
\hline
Graph-Based BERT \cite{saracino2023graph} & 96.45 & 95.79 & 94.09 & 93.91  \\
\hline
\end{tabular}
\end{table}

\begin{table}[htbp]
\centering
\caption{CLASSIFICATION REPORT of Transformers}
\label{tab:sample_table}
\begin{tabular}{|c|c|c|c|c|c|}
\hline
Category & Precision & Recall & F1   \\
\hline
adware\_bundler & 1.00 & 1.00 & 1.00     \\
\hline
adware\_domaiq & 1.00 & 1.00 & 1.00     \\
\hline
 emotet\_trj& 0.9980 & 0.9820  & 0.9899     \\
\hline
expiro\_virus &  0.9845  & 0.9969 &0.9906     \\
\hline
gandcrab\_trj  & 0.8571 & 0.9231  &  0.8889  \\
\hline
ekstak\_trj &  0.9167  & 1.0000  &  0.9565  \\
\hline

\end{tabular}
\end{table}

\begin{table}[htbp]
\centering
\caption{CLASSIFICATION REPORT of GNN}
\label{tab:sample_table}
\begin{tabular}{|c|c|c|c|c|c|}
\hline
Category & Precision & Recall & F1   \\
\hline
adware\_bundler & 0.9898 & 0.9803  & 0.9850   \\
\hline
adware\_domaiq & 0.9960  & 0.9917 & 0.9938     \\
\hline
 emotet\_trj&  0.9857 & 0.9963  & 0.9910     \\
\hline
expiro\_virus &   0.9908   &  0.9778 & 0.9843     \\
\hline
gandcrab\_trj  & 0.9803 & 0.9911   &  0.9857  \\
\hline
ekstak\_trj &  0.9700  & 0.9499  &  0.9599   \\
\hline

\end{tabular}
\end{table}

Table~I presents a comparison of different classification approaches
on our dataset. Because the source codes were unavailable, we
re-implemented the most representative methods from previous studies.
For the CFGExplainer \cite{herath2022cfgexplainer} baseline, we
reproduced the paper's ACFG setting by constructing one control-flow
graph per binary where nodes are basic blocks and edges are directed
control-flow. We computed the same 12-dimensional structural feature
vector per node. To enable batching, graphs were padded/truncated to
a fixed maximum size. For training a GCN, all graphs must have the
same maximum size N, so we used $N=7000$ nodes per sample and trained
with a 3-layer GCN followed by global pooling and a linear family
classifier. CFGExplainer achieved Precision, Recall, F1-score, and
Accuracy values of 72.2, 71.8, 70.24, and 83.7, respectively, which are
significantly lower than those obtained by our models. Furthermore,
CFGExplainer explicitly targets interpretability by outputting
important subgraphs and node orderings, but its explanations are
post-hoc and are mainly evaluated by whether the extracted subgraph
preserves the classifier's accuracy (fidelity), not whether the
highlighted blocks correspond to ground-truth malicious behavior.
Also, large subgraphs can still require exhaustive manual inspection.

For MalBERT \cite{rahali2021malbert}, we mapped each Windows binary into a BERT-style text document by disassembling basic blocks recovered from angr CFGFast, normalizing instruction strings to reduce address/register noise, and concatenating the normalized instructions into a single document per sample. The documents were tokenized with a BERT tokenizer (maximum sequence length 512 with truncation/padding) and trained using a standard HuggingFace training pipeline. Finally, for the graph-based BERT \cite{saracino2023graph} baseline, we constructed a call-graph view and applied a Kahn-style topological ordering to produce an ordered function sequence, which was then treated as text input to a BERT encoder. While the results of \cite{rahali2021malbert} and \cite{saracino2023graph} are notable, they fall short of the accuracy attained by our approach, demonstrating its superior performance. This underscores the effectiveness of studying behavioral patterns across diverse malware samples and validates our method as a reliable malware detection technique. Moreover, MalBERT and Graph-Based BERT primarily provide label predictions from sequence representations and do not connect decisions back to behavior subgraphs or code regions that implement malicious intent, leaving the classifier largely as a black box at the analyst level.

Beyond classification performance, we evaluate the interpretability of
our model by analyzing whether the model highlights semantically
meaningful malicious behaviors identified through manual analysis. Our
explainability assessment is conducted at the behavior level. For every
behavior, we assign (i) a binary maliciousness label \{0,1\} and (ii) a
continuous maliciousness score reflecting severity [0,1]. We applied
this evaluation to 10 malware samples spanning 9 malware families (one
family, LokiBot, is represented by two samples); their behavior
coverage results are reported in Table~IV. Samples were selected based
on the variation in Transformer classification performance observed in
Table~II (precision, recall, and F1), ranging from a very high-performing
sample, Emotet\_trj, to the family with the lowest performance,
gandcrab\_trj.

\begin{table}[t]
\centering
\caption{Malware Samples with Behavior Coverage}
\label{tab:behavior_coverage}
\begin{tabular}{l c c}
\hline
\textbf{Sample ID} & \textbf{Family} & \textbf{Behavior Coverage} \\
\hline
emotet\_trj/Virus1329     & Emotet        & 1.00 \\
gandcrab\_trj/Virus38    & GandCrab      & 0.90 \\
lokibot.trj/Virus348     & LokiBot       & 1.00 \\
lokibot.trj/Virus388     & LokiBot       & 1.00 \\
dridex\_trj/Virus3587    & Dridex        & 0.90 \\
ekstak\_trj/Virus14      & Ekstak        & 0.85 \\
trj.qakbot/Virus68       & Qakbot        & 1.00 \\
trj\_rozena/Virus90      & Rozena        & 0.85 \\
trj\_toolbar/Virus279    & Toolbar       & 0.87 \\
mydoom\_worm/Virus28     & Mydoom        & 0.80 \\
\hline
\end{tabular}
\end{table}

From the selected subset, we present case studies for Mydoom and Qakbot. In both cases, the model correctly predicts the malware family with high confidence and highlights behaviors consistent with known family-specific tactics. 
Table~\ref{tab:mydoom_top10_behaviors} presents the top-10 behaviors ranked by Transformer model contribution for the Mydoom/Virus28 sample. Each entry corresponds to a behavior whose embedding received high attention during the final classification decision, with contributions normalized such that the most influential behavior has a score of 1.0. The ranked behaviors are dominated by file I/O and network communication roles, which aligns with Mydoom’s known propagation strategy involving payload staging followed by outbound network activity. Specifically, repeated high-ranking fileio behaviors invoking SetFilePointer and WriteFile indicate structured file staging, while winsock -related calls such as send, gethostname, and closesocket reflect active network communication. Importantly, only semantically interpretable behavior categories are retained in the explanation, avoiding generic or ambiguous classes. This demonstrates that the model’s prediction is supported by coherent, malware-relevant behaviors rather than spurious or opaque features. 

\begin{table}[t]
\centering
\caption{Mydoom/Virus28: Top-10 behaviors ranked by Transformer model contribution.}
\label{tab:mydoom_top10_behaviors}

\footnotesize
\setlength{\tabcolsep}{4pt}

\begin{tabular}{c l l l l}
\hline
\textbf{Rank} & \textbf{Behavior Role(s)} & \textbf{Key API(s)} & \textbf{DLL} & \textbf{Contrib.} \\
\hline
1 & fileio & SetFilePointer & KERNEL32.DLL & 1.000 \\
2 & fileio & SetFilePointer & KERNEL32.DLL & 0.971 \\
3 & winsock & send & WS2\_32.dll & 0.954 \\
4 & winsock & send & WS2\_32.dll & 0.954 \\
5 & winsock & gethostname & WS2\_32.dll & 0.952 \\
6 & fileio & SetFilePointer & KERNEL32.DLL & 0.874 \\
7 & winsock & htons & WS2\_32.dll & 0.864 \\
8 & winsock & closesocket & WS2\_32.dll & 0.837 \\
9 & fileio & WriteFile & KERNEL32.DLL & 0.835 \\
10 & fileio & WriteFile & KERNEL32.DLL & 0.831 \\
\hline
\end{tabular}
\end{table}

\begin{table}[t]
\centering
\caption{Qakbot/Virus68: Top-10 behaviors ranked by Transformer model contribution.}
\label{tab:qakbot_top10_behaviors}
\vspace{-0.4em}
\setlength{\tabcolsep}{5pt}
\renewcommand{\arraystretch}{1.1}
\resizebox{\columnwidth}{!}{%
\begin{tabular}{c l l l c}
\hline
\textbf{Rank} & \textbf{Behavior Role(s)} & \textbf{Key API(s)} & \textbf{DLL} & \textbf{Contrib.} \\
\hline
1  & ipc\_pipe        & CreatePipe                 & KERNEL32.DLL & 1.000 \\
2  & ipc\_pipe        & CreatePipe                 & KERNEL32.DLL & 0.987 \\
3  & ipc\_pipe        & CreatePipe                 & KERNEL32.DLL & 0.939 \\
4  & ipc\_pipe        & CreatePipe                 & KERNEL32.DLL & 0.813 \\
5  & ipc\_pipe        & CreatePipe, memset         & KERNEL32.DLL & 0.805 \\
6  & registry\_policy & RegQueryValueExW           & KERNEL32.DLL & 0.751 \\
7  & registry\_policy & RegQueryValueExW           & KERNEL32.DLL & 0.748 \\
8  & registry\_policy & RegDeleteValueW            & KERNEL32.DLL & 0.732 \\
9  & registry\_policy & RegQueryValueExW           & KERNEL32.DLL & 0.698 \\
10 & registry\_policy & RegDeleteValueW            & KERNEL32.DLL & 0.641 \\
\hline
\end{tabular}}
\vspace{-0.6em}
\end{table}


Table~VI presents the top-10 behaviors ranked by Transformer model
contribution for the Qakbot/Virus68 sample. The model's explanation is
dominated by inter-process communication and registry manipulation
behaviors. The highest-ranked behaviors consistently involve named
pipe creation via CreatePipe, indicating IPC-based coordination or
command handling. Complementary registry access and deletion operations
(RegQueryValueExW, RegDeleteValueW) reflect persistence and
configuration management mechanisms characteristic of Qakbot malware.
Repeated entries in the top-ranked behaviors reflect multiple independent behavior subgraphs exhibiting the same functional role and API usage, indicating consistent emphasis on family-specific malicious routines rather
than reliance on a single artifact.

\section{Conclusion}
\label{sec:conclusion}

This paper presented a behavior-centric malware analysis framework that moves beyond binary-level classification by explicitly linking system-level behaviors to their originating basic-block regions. The pipeline anchors the representation around security-sensitive API sinks and reconstructs the execution context that leads to those sinks as basic-block–level behavior subgraphs. We then incorporate manual labeling through a Malicious DNA reference, assigning normalized threat scores that distinguish behavior-critical logic from surrounding boilerplate. These score-aware representations are consumed by two complementary models: a Behavior Transformer that captures ordered behavior narratives and a score-aware GNN that captures structural dependencies within behavior graphs. Evaluated on 43 malware families, both branches achieve strong performance, and decision-level late fusion yields the most consistent results, reaching 99.87\% across accuracy, precision, recall, and F1. Beyond classification, the framework introduces a behavior-coverage metric that measures whether the model’s highest-importance regions overlap with manually identified malicious behaviors, on representative samples, coverage remains high and supports qualitative case-study analysis.

\bibliographystyle{IEEEtran}
\bibliography{sample}
\end{document}